\documentclass[aps,twocolumn,showpacs,amssymb,groupedaddress,preprintnumbers,showpacs,nofootinbib]{revtex4-1}
\usepackage{graphicx}
\usepackage{amsmath}
\usepackage{amsfonts}
\usepackage{amssymb}
\usepackage{epsfig}
\newcommand{\be}{\begin{equation}}
\newcommand{\ee}{\end{equation}}

\newcommand{\no}{\nonumber\\}
\newcommand{\ba}{\begin{eqnarray}}
\newcommand{\ea}{\end{eqnarray}}
\newcommand{\kk}{|\vec k|}
\newcommand{\kkk}{|\vec k'|}
\newcommand{\hh}{\eta_0}

\begin{document}
\preprint{ICCUB-11-146~~~~ UB-ECM-PF-11-52}
\title{Photon propagation in a cold axion background with and without magnetic field}
\author{D. Espriu}
\author{A. Renau}
\affiliation{Departament d'Estructura i Constituents
de la Mat\`eria and Institut de Ci\`encies del Cosmos (ICCUB) ,
Universitat de Barcelona, Mart\'\i \ i Franqu\`es 1, 08028 Barcelona, Spain}


\begin{abstract}
A cold relic axion condensate resulting from vacuum misalignment
in the early universe oscillates with a frequency $m$, where $m$ is the
axion mass.
We determine the properties of photons propagating in a simplified version of such
a background where the sinusoidal variation is replaced by a square wave profile.
We prove that previous results that indicated that charged particles moving fast in
such a background radiate, originally derived assuming that all momenta involved
were much larger than $m$, hold for long wavelengths too. We also analyze in detail
how the introduction of a magnetic field changes the properties of
photon propagation in such a medium. We briefly comment on possible astrophysical implications
of these results.
\end{abstract}
\maketitle


\section{Introduction}

Axions, originally introduced to solve the strong CP problem\cite{PQ},
are to this date a viable candidate to constitute the dark matter of the
universe\cite{darkmatter}. Their contribution to the mass density results from the energy stored
in the collective oscillations around the minimum of the axion potential
\be
a(t)=a_0\cos(mt),
\ee
with a frequency that is given by the axion mass $m$. We know that this mass must be somewhere in
the range\cite{other}
\be
1 \text{ eV} > m > 10^{-6} \text{ eV}.
\ee
The coupling of axions to photons takes place through the
universal term\footnote{This term is often written as
$\displaystyle \mathcal L_{A\gamma\gamma}= \frac{G_{A\gamma\gamma}}4F_{\mu\nu}\tilde F^{\mu\nu}\phi_A$,
where $\phi_A$ is the axion field\cite{pdg}. Both $G_{A\gamma\gamma}$ and $g_{a\gamma\gamma}$ are used interchangeably in
the axion literature as coupling constants having dimensions $E^{-1}$. The constant $g_{a\gamma\gamma}$ used here
is howeever dimensionless
and it should not be confused with the latter.}
\be\label{agg}
\mathcal L_{a\gamma\gamma}=g_{a\gamma\gamma}\frac\alpha{2\pi}\frac a{f_a}F_{\mu\nu}\tilde F^{\mu\nu},
\ee
where $\tilde F^{\mu\nu}=\frac12\epsilon^{\mu\nu\alpha\beta}F_{\alpha\beta}$ is the
dual electromagnetic tensor. The dimensionful quantity $f_a$ is the axion decay constant -- the
equivalent of $f_\pi$ as axions are assumed to be the pseudo Goldstone bosons associated to the
breaking of the Peccei-Quinn symmetry $U_{PQ}(1)$\cite{PQ}. On $f_a$ we have a range of bounds:
$f_a > 10^4$ GeV coming from direct experimental searches of axions coupling directly to matter\cite{searches};
$f_a > 10^7 $ GeV from (somewhat weaker) astrophysical constraints\cite{mass}, largely mass independent;
or $f_a > 10^7$ for $0.02~{\rm eV}< m < 0.4~{\rm eV}$ coming from the phase II of the
CAST experiment\cite{cast}. For some reviews of the experimental/observational search for axions
see \cite{other}.

The constant $g_{a\gamma\gamma}$ is model dependent, but it is typically of order 1
in most axion models\cite{models}. The axion, being a pseudo Goldstone boson,
satisfies the relation $f_a m \simeq {\rm constant}\simeq f_\pi m_\pi $, thus constraining the basic parameters
of the theory. However, the results presented below apply also to other light pseudoscalar particles, sometimes
termed axion-like particles (ALP). The coupling between ALP and photons could in principle be stronger,
since it is not related to their mass.

Integrating by parts, we can write the term coupling axions or ALP to photons like
\be\label{eta}
\mathcal L_{a\gamma\gamma}=\frac12\eta_\mu A_\nu\tilde F^{\mu\nu},
\ee
with
\be
\eta_\mu=\eta(t)\delta_{\mu0}~,\quad\eta(t)= \eta_0 \sin m t.
\ee
The Lagrangian for a photon in the cold axion background is then
\be
\mathcal L=-\frac14F^{\mu\nu}F_{\mu\nu}+\frac12\eta_\mu A_\nu\tilde F^{\mu\nu},
\ee
and the relevant quantity to determine the physical effect of this coupling is
\be
\eta_0= 2g_{a\gamma\gamma}\frac{\alpha}{\pi}\frac{ a_0 m}{f_a}.
\ee

Now we can proceed to quantizing the photon field in such a background. This has been
previously done in \cite{sasha} in the case where $\eta(t)$ is assumed to be a constant,
$\eta(t)=\eta_0$. It was found that in this case the two physical photon polarizations get their dispersion
relations modified in the following way
\be
\omega_\pm = \sqrt{\vec k^2\pm \hh \vert \vec k\vert}.
\ee
As a consequence processes that are forbidden on Lorentz-invariance grounds such
as $ \gamma \to e^+ e^- $  or $e\to e \gamma$ have a non-vanishing probability
if certain kinematical constraints are fulfilled. The interested reader can see
\cite{cosmicrays} for possible observable consequences. If measured, these effects
would constitute \emph{prima facie} evidence that not only axions or ALP exist but they do constitute
the primary ingredient of the dark matter of the universe.

It was argued in \cite{cosmicrays,ourfirst}
that taking $\eta(t)$ as a constant was a good approximation if the momenta of all
particles involved in the process were larger than $m$, the period of
oscillations. However, if the wavelength of some of the particles are comparable or lower than the
period of oscillation one must necessarily deal with a time-dependent external potential.
Thus it seems to us quite important to establish
the basic principles of photon propagation in a time dependent axion background.
For this reason in this paper we solve the problem of photon propagation in
an oscillatory, but spatially constant, axion background exactly. We shall also
include an external magnetic field to see how the combined effect modifies
the properties of photons moving in such an environment. We will discuss
at the end of the paper some possible physical consequences.

To keep the paper technically simple we have approximated the sinusoidal time
dependence of the background by a square wave with the same period. A sinusoidal
wave involves Mathieu special functions complicating the calculation
enormously.  We base this approximation on the similarity of the present
effect with the emergence of the band structure in periodic potentials\cite{kp}, exchanging
time and space, and momenta and energies. It is well known in solid state physics that even such
a simple model fully captures the esentials of metallic conductors and semi-conductors.
Therefore we firmly believe that the physics of the problem being discussed remains unaltered by
our technical simplification.

\section{Solving for the eigenmodes and eigenvalues}

We introduce a Fourier transform with respect to the spatial coordinates only
and write the photon field as
\be\label{fourier}
A_\mu(t,\vec x)=\int \frac{d^3k}{(2\pi)^3}e^{i\vec k\cdot\vec x}\hat A_\mu(t,\vec k).
\ee
The equation for $\hat A_\nu(t,\vec k)$ is
\be\label{eq}
\left[g^{\mu\nu}(\partial_t^2+\vec k^2)-i\epsilon^{\mu\nu\alpha\beta}\eta_\alpha k_\beta\right]\hat A_\nu(t,\vec k)=0.
\ee
We now define
\be
S^\nu_{~\lambda}=\epsilon^{\mu\nu\alpha\beta}\eta_\alpha k_\beta\epsilon_{\mu\lambda\rho\sigma}\eta^\rho k^\sigma,
\ee
which can also be written as
\ba
S^{\mu\nu}=&&\left[(\eta\cdot k)^2-\eta^2k^2\right]g^{\mu\nu}+k^2\eta^\mu\eta^\nu\\&&+\eta^2k^\mu k^\nu-
(\eta\cdot k)(\eta^\mu k^\nu+\eta^\nu k^\mu),
\ea
and
\be\label{proj}
P_\pm^{\mu\nu}=\frac{S^{\mu\nu}}S\mp\frac i{\sqrt{2S}}\epsilon^{\mu\nu\alpha\beta}\eta_\alpha k_\beta~,\quad S=
S^\mu_{~\mu}=2\eta^2\vec k^2.
\ee
The properties of these quantities are discussed in \cite{sasha}. Note that the time dependence (due to $\eta(t)$)
in $P_\pm^{\mu\nu}$ cancels. With the help of these projectors we can write \eqref{eq} as
\be\label{eq2}
\left[g^{\mu\nu}(\partial_t^2+\vec k^2)+\sqrt{\frac S2}\left(P_+^{\mu\nu}-
P_-^{\mu\nu}\right)\right]\hat A_\nu(t,\vec k)=0.
\ee
To solve the equations of motion we introduce the polarization vectors defined in \cite{sasha}
and write\footnote{When $\eta_\mu$ only has a temporal component, these polarization vectors actually reduce to the usual ones.}
\be
\hat A_\nu(t,\vec k)=\sum_{\lambda=+,-}f_\lambda(t)\varepsilon_\nu(\vec k,\lambda).
\ee
These vectors satisfy
\be
P^{\mu\nu}_\pm\varepsilon_\nu(\vec k,\pm)=\varepsilon^\mu(\vec k,\pm),
\quad P^{\mu\nu}_\pm\varepsilon_\nu(\vec k,\mp)=0
\ee
and do not depend on $t$, so
\be
\left[\partial_t^2+\vec k^2\pm\eta(t)\kk\right]f_\pm(t)=0.
\ee
As mentioned we will approximate the sine function in $\eta(t)$ by a square wave function:
\be\label{square}
\eta(t)=\left\{\begin{array}{cc}
                 +\hh & 2nT<t<(2n+1)T \\
                 -\hh & (2n+1)T<t<2nT
               \end{array}
\right..
\ee
The relevant parameters are
\be
\eta_0=2g_{a\gamma\gamma}\frac{\alpha}{\pi}\frac{ a_0 m}{f_a},\qquad T=\frac\pi{m}.
\ee
There is an equation for each polarization. However, they are related.
To recover one from the other we can just make the replacement $\hh\rightarrow-\hh$. Also,
because $\eta(t)$ changes sign after a time $T$ in the square wave approximation one solution
is a time-shifted copy of the other: $f_-(t)=f_+(t+T)$.
In what follows we will work in the case $\lambda=+$. It is obvious that the conclusions also apply to the
other physical polarization, $\lambda=-.$

Since $\eta(t)$ is piecewise-defined, we will solve the equation in two regions:\\
-- Region 1:  $0<t<T$, $\eta(t)=\hh$
\be
\frac{d^2f_1(t)}{dt^2}+(\vec k^2+\hh \kk)f_1(t)=0,
\ee
\be\label{alpha}
f_1(t)=A'e^{i\alpha t}+Ae^{-i\alpha t}~,\quad\alpha^2=\vec k^2+\hh\kk.
\ee
-- Region 2: $-T<t<0$, $\eta(t)=-\hh$
\be
\frac{d^2f_2(t)}{dt^2}+(\vec k^2-\hh\kk)f_2(t)=0,
\ee
\be\label{beta}
f_2(t)=B'e^{i\beta t}+Be^{-i\beta t}~,\quad\beta^2=\vec k^2-\hh\kk.
\ee
We impose that both functions coincide at $t=0$ and we do the same for their derivatives
\be
f_1(0)=f_2(0), \qquad f^\prime_1(0)=f^\prime_2 (0).
\ee
We now write $f(t)=e^{-i\omega t}g(t)$ and demand that $g(t)$ have the same periodicity as $\eta(t)$
\ba
&g_1(t)=e^{i\omega t}f_1(t)=A'e^{i(\omega+\alpha) t}+Ae^{i(\omega-\alpha) t},\cr
&g_2(t)=e^{i\omega t}f_2(t)=B'e^{i(\omega+\beta) t}+Be^{i(\omega-\beta) t},\cr
&g_1(T)=g_2(-T),\quad g_1'(T)=g_2'(-T).
\ea
For these conditions to be fulfilled, the coefficients have to solve the linear system
\begin{align}
A'+A&=B'+B\no
\alpha A'-\alpha A&=\beta B'-\beta B\no
\nonumber e^{i(\omega+\alpha)T}A'+e^{i(\omega-\alpha)T}A&=e^{-i(\omega+\beta)T}B'+e^{-i(\omega-\beta)T}B
\end{align}
\vspace{-2.5em}
\begin{align}
\!\!\!(\omega+\alpha)&e^{i(\omega+\alpha)T}A'+(\omega-\alpha)e^{i(\omega-\alpha)T}A\no
&~~~=(\omega+\beta)e^{-i(\omega+\beta)T}B'+(\omega-\beta)e^{-i(\omega-\beta)T}
.\end{align}
The linear system can be expressed as
\be
{\hat M}
\left(
  \begin{array}{c}
    A'  \\
    A \\
    B'  \\
    B \\
  \end{array}
\right)=
\left(
  \begin{array}{c}
    0 \\
    0 \\
    0 \\
    0 \\
  \end{array}
\right)
,\ee
with
\be
{\hat M^T}=\left(
  \begin{array}{cccc}
    1 & \alpha & e^{i(\omega+\alpha)T} & (\omega+\alpha)e^{i(\omega+\alpha)T} \\
    1 & -\alpha & e^{i(\omega-\alpha)T} & (\omega-\alpha)e^{i(\omega-\alpha)T} \\
    -1 & -\beta & -e^{-i(\omega+\beta)T} & -(\omega+\beta)e^{-i(\omega+\beta)T} \\
    -1 & \beta & -e^{-i(\omega-\beta)T} & -(\omega-\beta)e^{-i(\omega-\beta)T} \\
  \end{array}
\right).
\ee

The problem being discussed here is formally similar to the solution of the Kronig-Penney\cite{kp}  one-dimensional
periodic potential, except the periodicity is now in time rather than in space.

In order to find a non-trivial solution one has to demand the condition of
vanishing determinant of $\hat M$, which is
\be
\cos(2\omega T)=\cos(\alpha T)\cos(\beta T)-\frac{\alpha^2+\beta^2}{2\alpha\beta}\sin(\alpha T)\sin(\beta T),
\label{kp}
\ee
with $\alpha$ and $\beta$ given by \eqref{alpha} and \eqref{beta} respectively.
In order to get analytical expressions we will work in
the limit of long wavelengths $\kk T\ll 1$, which is just the one
that is potentially problematic as discussed in the introduction.
Expanding both sides:
\be
\omega^2-\frac13\omega^4T^2+...=\vec k^2-\left(\frac13\vec k^4-\frac1{12}\hh^2\vec k^2\right)T^2+...,
\ee
which means
\be
\omega^2\approx\left(1+\frac{\hh^2T^2}{12}\right)\vec k^2.
\ee
If the determinant vanishes the system to solve is
\be
\left(
  \begin{array}{ccc}
   1  & 1 & -1 \\
    0  & 1  & -\frac12(1-\frac\beta\alpha) \\
     0 & 0 & 1 \\
  \end{array}
\right)\left(
         \begin{array}{c}
           A' \\
           A \\
           B' \\
         \end{array}
       \right)=\left(
                 \begin{array}{c}
                   1 \\
                   \frac12(1+\frac\beta\alpha) \\
                   h(\alpha,\beta,T) \\
                 \end{array}
               \right)B,
\ee
where
\be
h(\alpha,\beta,T)=-\frac{\alpha-\beta}{\alpha+\beta}\frac{e^{i\alpha T}-e^{-i2\omega T}e^{i\beta T}}{e^{i\alpha T}-e^{-i2\omega T}e^{-i\beta T}},
\ee
leading to
\ba
\frac{A'}B&=&\left[1-\frac{\alpha-\beta}{\alpha+\beta}\frac{e^{i\alpha T}e^{i2\omega T}-e^{i\beta T}}{e^{i\alpha T}e^{i2\omega T}-e^{-i\beta T}}\right.\cr&&\left. - \frac12(1+\frac\beta\alpha)+\frac12(1-\frac\beta\alpha)\frac{\alpha-\beta}{\alpha+\beta}\frac{e^{i\alpha T}e^{i2\omega T}-e^{i\beta T}}{e^{i\alpha T}e^{i2\omega T}-e^{-i\beta T}}\right]\cr
\frac AB&=&\left[\frac12(1+\frac\beta\alpha)-\frac12(1-\frac\beta\alpha)\frac{\alpha-\beta}{\alpha+\beta}\frac{e^{i\alpha T}e^{i2\omega T}-e^{i\beta T}}{e^{i\alpha T}e^{i2\omega T}-e^{-i\beta T}}\right]\cr
\frac{B'}B&=&\left[-\frac{\alpha-\beta}{\alpha+\beta}\frac{e^{i\alpha T}e^{i2\omega T}-e^{i\beta T}}{e^{i\alpha T}e^{i2\omega T}-e^{-i\beta T}}\right].
\ea

In the limit $\hh\ll\kk$, $\kk T\ll1$,
\be\label{limit}
\frac{A'}B\approx-\frac{B'}B\approx\frac14\frac{\hh} \kk, \quad \frac{A}B\approx1-\frac{\hh}{2\kk}
.\ee
Finally, imposing the usual normalization,
\be
\displaystyle\int f_k(t)f_{k'}^*(t)=2\pi\delta(\kk-\kkk),
\ee
we get
\ba
B&=&\left[\frac{\sqrt{\vec k^2+\hh \kk}}{2\kk+\hh}\left(\left|\frac AB\right|^2+
\left|\frac {A'}B\right|^2\right)\right.\cr&&\left.+\frac{\sqrt{\vec k^2-
\hh \kk}}{2\kk-\hh}\left(1+\left|\frac {B'}B\right|^2\right)\right]^{-1/2}\approx\left(1+\frac{\hh}{4\kk}\right).\no
\ea
This completes the determination of the eigenvectors.

\subsection{Exact determination of the eigenvalues}

\begin{widetext}
\begin{figure}[h!]
\begin{center}
\includegraphics[scale=0.45,trim=-5.5cm 0 0 0]{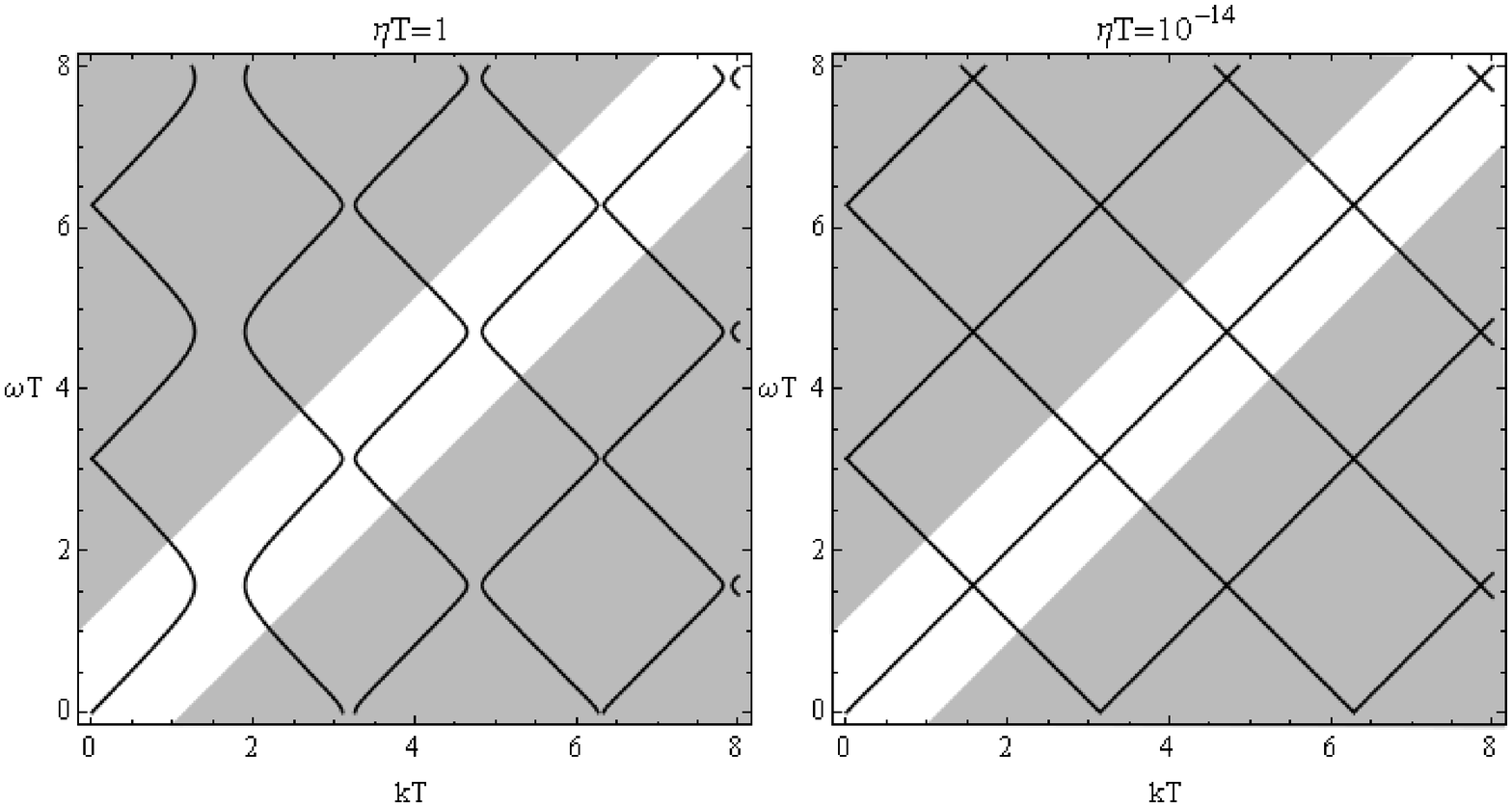}
\end{center}
\parbox{15cm}{\caption{Plots of the solution for $\eta_0 T=1$ and $\eta_0 T=10^{-14}$. In the $\eta_0 T\to 0$ limit the solutions
correspond to the straight lines $\omega\sim \kk$ (plus their periodic repetitions). Small gaps develop but they
become only physically significant when $\eta_0 T = \mathcal O(1)$. The physical region corresponds to the white area, the gray areas
are just periodic repetitions.\label{gaps}}}
\end{figure}
\end{widetext}

We can also solve \eqref{kp} exactly, without having to assume the long-wavelength limit as above, but
this can be done only numerically. The solution only depends on $\eta_0$ and $m$ through the dimensionless
combination $\eta_0 T$.
There are values of $k$ for which there is no solution, as seen in figure \ref{gaps}. However, these gaps get
narrower when the product $\eta_0 T$ decreases. In practice, the largest possible physical value
for this quantity is $\eta_0 T=10^{-14}$ and then the
gaps are practically nonexistent and certainly totally irrelevant for the purposes of the present paper.

It is interesting to investigate whether  {\it complex} solutions exist for $\omega$ in the forbidden narrow bands.
We note that the R.H.S. of the equation (\ref{kp}) is necessarily real, thus $\omega$ must necessarily be purely
real or purely imaginary. In the latter case the L.H.S. is replaced by a $\cosh$ having as argument the imaginary
part of $2\omega T$. For this to have a solution, the R.H.S. must be positive and larger than one. Inspection of
this term reveals that it is larger that one in the forbidden zones but actually alternates sign. Therefore not even
an imaginary solution exists for the first, third,... forbidden regions.

\subsection{Calculation of the transition $e\to e \gamma$}

In order to make the photon field hermitian, we add \eqref{fourier} and its conjugate. Introducing creation
and annihilation operators for each one of the proper modes we get (both polarizations are included)
\ba
A_\mu(t,\vec x)&=&\int \frac{d^3k}{(2\pi)^3}
\sum_\lambda\left[a(\vec k,\lambda)g(t,\vec k,\lambda)\varepsilon_\mu(\vec k,\lambda)e^{-ikx}\right.\cr&&\left.
+a^\dag(\vec k,\lambda)g^*(t,\vec k,\lambda)\varepsilon^*_\mu(\vec k,\lambda)e^{ikx}\right],\ea
where $kx\equiv\omega t-\vec k\cdot\vec x$. Now we want to compute $\langle f|S|i\rangle$ for an
initial state of one electron of momentum $p$
and a final state of an electron of momentum $q$ and a photon of momentum $k=p-q$.

\ba
\langle f|S|i\rangle&=& ie\varepsilon^*_\mu(\vec k,\lambda)\bar u_q\gamma^\mu u_p(2\pi)^3\delta^{(3)}
(\vec k+\vec q-\vec p)\cr&&\times\int dt g^*(t,\vec k,\lambda)e^{i(\omega+E_q-E_p)t}
\ea
If we take $\eta(t)$ constant, $g(t,\vec k,\lambda)=1$ and we have
\be
\langle f|S|i\rangle=ie\varepsilon^*_\mu(\vec k,\lambda)\bar u_q\gamma^\mu u_p(2\pi)^4\delta^{(4)}(k+q-p).
\ee
In the square wave approximation \eqref{square}, the time integration yields

\begin{widetext}
\ba
\langle f|S|i\rangle
&=&ie\bar u_q\gamma^\mu u_p\varepsilon^*_\mu(\vec k,\lambda)(2\pi)^3\delta(\vec k+\vec q-\vec p)\pi
\left\{A\delta(\alpha+E_q-E_p)\right.\cr&&\left.
+B \delta(\beta+E_q-E_p)+ A'
\delta(-\alpha+E_q-E_p)-B'\delta(-\beta+E_q-E_p)\right\}\label{AB}\\
&\approx&ie\bar u_q\gamma^\mu u_p\varepsilon^*_\mu(\vec k,\lambda)\left
(1+\frac{\hh}{4\kk}\right)(2\pi)^3\delta(\vec k+\vec q-\vec p)\pi\left\{\left(1-\frac{\hh}{2\kk}\right)\delta(\alpha+E_q-E_p)\right.\cr&&\left.
+ \delta(\beta+E_q-E_p)+ \frac{\hh}{4\kk}
\left[\delta(-\alpha+E_q-E_p)+\delta(-\beta+E_q-E_p)\right]\right\}.
\ea
\end{widetext}
Equation \eqref{AB} holds for any value of $\kk$. The $\approx$ symbol indicates the use of \eqref{limit}.
It turns out that at the leading order in the $\hh$ expansion this expression agrees exactly
with the one obtained in \cite{ourfirst} assuming that $\eta(t)$ was constant except for the fact
that for each value of the polarization only one of the two delta functions
that are not suppressed by terms of the form $\hh/\kk$ can be simultaneously satisfied; namely
the one that implies that $\alpha$ or $\beta$ equals $\sqrt{\vec k^2-|\hh|\kk}$, contributing with a factor $1/2$ with respect
to what is found for constant $\eta$ to the amplitude. Thus in the transition
reduced matrix element $i\mathcal M$ one gets for each polarization exactly one-half of what
is obtained if $\eta(t)$ is constant. But in the present case both polarizations contribute so finally we get
$(1/2)^2+(1/2)^2=1/2$ of the result obtained with constant $\eta(t)$.

As a consequence the predictions concerning the radiation yield of a high energy charged particle
propagating in the cold axion background\cite{cosmicrays} are confirmed.

\section{Propagation in a magnetic field}

We will now compute the propagator of the photon field with two backgrounds: a cold axion background and
a constant magnetic field. To do so, we take \eqref{agg} and write the axion and photon fields as a
background term plus a dynamical field. We get two relevant terms
\be\label{vertexs}
\mathcal L_{a\gamma\gamma}\rightarrow\frac12\epsilon^{\mu\nu\alpha\beta}\eta_\mu A_\nu\partial_\alpha A_\beta
+\frac{2g_{a\gamma\gamma}\alpha}{\pi f_a}a\partial_\mu A_\nu \tilde F^{\mu\nu},
\ee
where $a$ is the axion field, $A_\mu$ is the photon field and $\tilde F^{\mu\nu}$ corresponds
to a magnetic field: $\tilde F^{0i}=B^i$, $\tilde F^{ij}=0$.
The first term is just \eqref{eta}. Here we will take $\eta(t)$ to be constant; therefore the results that follow
are valid only if the distance travelled by the photon, $l$, verifies $l< 2\pi/m$.

\begin{figure}[h]
\center
\includegraphics[scale=0.4]{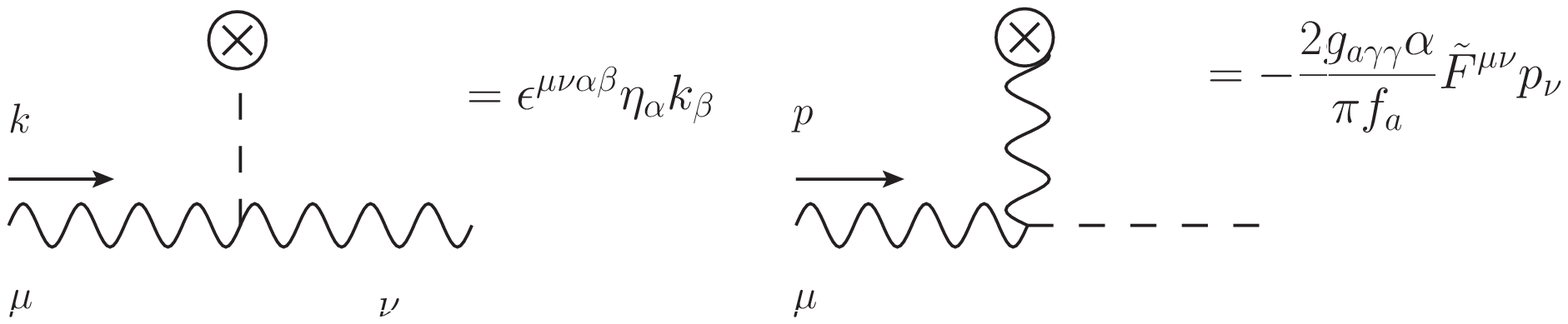}
\caption{The two relevant vertices. The corresponding Feynman rules are shown.}
\label{vertex}
\end{figure}

\begin{figure}[h]
\center
\includegraphics[scale=0.38]{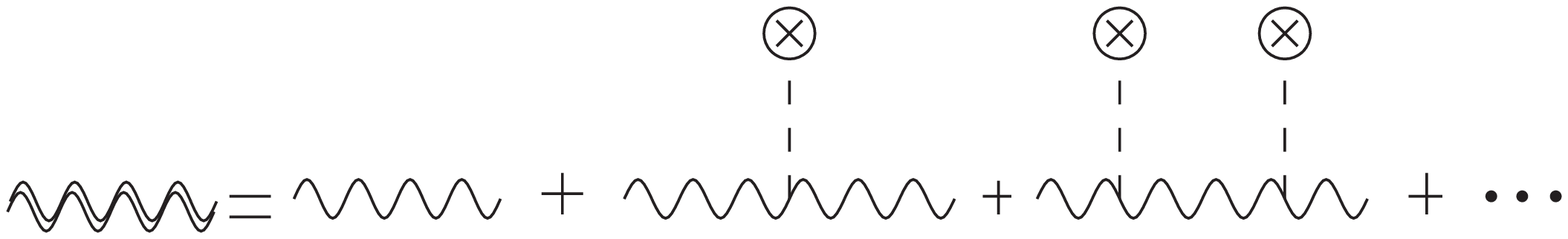}
\caption{Propagator in the axion background.}
\label{prop1}
\end{figure}

The vertices and Feynman rules corresponding to these terms are shown in figure \ref{vertex}.
With the first vertex we can compute the propagator in an axion background, see figure \ref{prop1}.
The successive interactions with the axion background can be summed up and the result
is the propagator
\be
D^{\mu\nu}=-i\left(\frac{g^{\mu\nu}-X^{\mu\nu}}{k^2
}+\frac{P_+^{\mu\nu}}{k^2-\hh \kk
}
+\frac{P_-^{\mu\nu}}{k^2+\hh\kk
}\right).\label{photonprop}
\ee
The physical polarizations, projected out by $P_\pm^{\mu\nu}$, exhibit poles at
$\omega^2=\vec k^2 \pm \eta_0\vert \vec k\vert$ as expected. The projectors are
defined in \eqref{proj} and $\displaystyle X^{\mu\nu}=\frac{S^{\mu\nu}}{\hh^2\kk^2}$.
Of course the same result can be obtained by direct inversion of the photon equation
of motion (\ref{eq}).

We now compute the propagator in the presence of a magnetic field, using the second term in \eqref{vertexs}.
In order to do that we use the propagator just found, represented by a double-wavy line and include the
interactions with the external magnetic field. The dashed line corresponds to the axion propagator.
\begin{figure}[h]
\center
\includegraphics[scale=0.38]{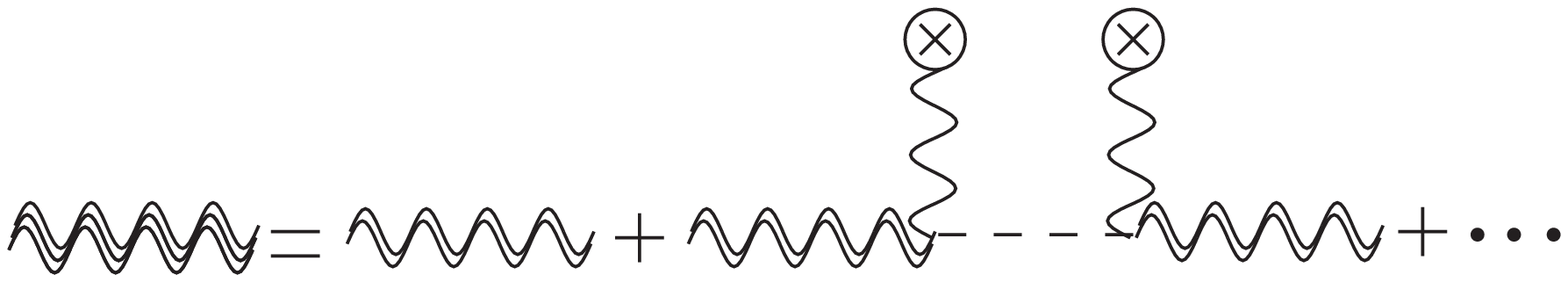}
\caption{Full propagator after resummation of the interactions with the external $\vec B$ field.}
\label{prop2}
\end{figure}
Summing all the diagrams we get
\be
\mathcal D_{\mu\nu}=D_{\mu\nu}+f_\mu h_\nu\frac{-ig^2}{k^2-m^2+ig^2K
},
\ee
where
\ba
&f_\mu=D_{\mu\alpha}\tilde F^{\alpha\lambda}k_\lambda,\quad &h_\nu=\tilde F^{\sigma\phi}k_\phi D_{\sigma\nu},\\
&\displaystyle g=\frac{2g_{a\gamma\gamma}\alpha}{\pi f_a}, \quad& K=\tilde F^{\beta\rho}k_\rho D_{\beta\gamma}\tilde F^{\gamma\xi}k_\xi.
\ea
In order to simplify the result we shall assume that $\vec k \cdot \vec B=0$, which may correspond
to an experimentally relevant situation. Then we get
\be
f_\mu=ik_0g^i_\mu\frac{k^2B_i-i\hh(\vec B\times\vec k)_i}{(k^2-\hh\kk)(k^2+\hh\kk)}
\ee
\be
h_\nu=ik_0g^j_\nu\frac{k^2B_j+i\hh(\vec B\times\vec k)_j}{(k^2-\hh\kk)(k^2+\hh\kk)}
\ee
\be
K=ik_0^2\vec B^2\frac{k^2}{(k^2-\hh\kk)(k^2+\hh\kk)},
\ee
and finally,  defining $\vec b\equiv g\vec B$,
\ba\label{propagator}
\mathcal D_{\mu\nu}&=&D_{\mu\nu}+ik_0^2g_\mu^j g_\nu^l\left\{\frac{b_jb_l
 }{(k^4-\hh^2\vec k^2)(k^2-m^2)-k_0^2 k^2\vec b^2 }\right.\cr
&&\left.+\frac{i\hh k^2\left[b_j(\vec b\times \vec k)_l-b_l(\vec b\times \vec k)_j\right]-
\hh^2 \vec b^2\vec k^2 X_{jl}}{(k^4-\hh^2\vec k^2)\left[(k^4-\hh^2\vec k^2)(k^2-m^2)-k_0^2 k^2\vec b^2 \right]}\right\}.\no
\ea

\subsection{Particular case: no axion background}

As a relevant particular case we now set $\eta_0=0$ in the previous expression, i.e.
we consider only the influence of the magnetic background,  and get
\be
\mathcal D_{\mu\nu}=D_{\mu\nu}+ik_0^2g_\mu^j g_\nu^l
\frac{ b_jb_l}{k^4(k^2-m^2)-k_0^2 k^2 \vec b^2},\label{propmagnetic}
\ee
where now $D_{\mu\nu}$ stands for the usual photon propagator, obtained after
setting $\eta_0=0$ in (\ref{photonprop}).

This propagator has poles when $k_0^2=\vec k^2$ and also for
\be
k_0^2=\frac12 \left(2\vec k^2+m^2+\vec b^2\pm\sqrt{m^4+\vec b^4+2m^2\vec b^2+4\vec b^2\vec k^2}\right)
.\ee
If we assume that $\vert \vec b\vert $ is a small parameter and expand in powers of it,
these poles in the frequency plane lie at
\ba\label{poles}
k_0^2&\simeq &\vec k^2\left( 1 +\frac{\vec b^2}{m^2}\right)+m^2+ \vec b^2, \cr
k_0^2&\simeq &\vec k^2\left( 1-\frac{\vec b^2}{m^2}\right).
\ea
Physically this pole structure corresponds to the perpendicular polarization vector $\epsilon_\perp$
propagating unchanged, while the parallel polarization $\epsilon_\parallel$  and the would-be
longitudinal polarization change their propagation\footnote{The labels $\perp$ and $\parallel$ refer to
directions perpendicular and parallel to the electric field, respectively, in the plane orthogonal
to the propagation.}.

For completeness we give the full propagator without the assumption $\vec k \cdot \vec B=0$
\be
\mathcal D_{\mu\nu}
=D_{\mu\nu}+f_\mu h_\nu\frac{i}{k^2-m^2+K}
\ee
where $D_{\mu\nu}=-ig_{\mu\nu}/k^2$ is the usual photon propagator and
\be
f_\mu=h_\mu=\frac 1{k^2}(g_{\mu0}\vec b\cdot\vec k+g_{\mu j}b^jk_0)
\ee
\be
K=\frac 1{k^2}\left[(\vec b\cdot\vec k)^2-\vec b^2k_0^2\right]
\ee

Let us now restore the condition $\vec k \cdot \vec B =0$ that is helpful in simplifying the formulae.
In order to write the propagator in a more compact form we introduce a four-vector $b^\mu=(0,\vec b= g\vec B)$
\be
\mathcal D_{\mu\nu}(k)=\frac{-i g_{\mu\nu}}{k^2}+\frac{ik_0^2b_\mu b_\nu}{k^2[k^2(k^2-m^2)-k_0^2\vec b^2]}.
\ee
Note the rather involved structure of the dispersion relation implied by (\ref{propmagnetic}).
We consider the propagation of plane waves of well defined frequency $\omega=k_0$ and moving in the $\hat x$ direction.
The Fourier transform with respect to the spatial component will describe the space evolution of a photon state emitted
at $x=0$ with polarization given by the vector $\epsilon_0$. We decompose
\be
\frac1{k^2[k^2(k^2-m^2)-k_0^2\vec b^2]}=\frac{A}{\vec k^2-k_0^2}+\frac{B}{\vec k^2-F}+\frac{C}{\vec k^2-G},
\ee
where $F$ and $G$ are the roots of the denominator
\ba
G&=&k_0^2-\frac{m^2}2-\frac12\sqrt{m^4+4k_0^2\vec b^2}\approx\left(1-\frac{\vec b^2}{m^2}\right)k_0^2-m^2,\no
F&=&k_0^2-\frac{m^2}2+\frac12\sqrt{m^4+4k_0^2\vec b^2}\approx\left(1+\frac{\vec b^2}{m^2}\right)k_0^2,
\ea
in agreement with \eqref{poles}, and
\ba
A&=&\frac1{k_0^2-F}\frac1{k_0^2-G}\approx-\frac1{k_0^2\vec b^2}\\
B&=&-\frac1{k_0^2-F}\frac1{F-G}\approx\frac1{k_0^2\vec b^2}\\
C&=&\frac1{k_0^2-G}\frac1{F-G}\approx\frac1{m^4+3k_0^2\vec b^2}\approx\frac1{m^4}.
\ea
Even for the largest magnetic fields conceivable the product $b= g B$ is rather small
compared to the range of acceptable values of the axion mass and it appears
justified to neglect $C$. The space Fourier transform of the propagator is then
\ba
\mathcal D_{\mu\nu}(k_0,x)&=&-\frac{ g_{\mu\nu}}{2k_0}e^{ik_0|x|}-
\frac{k_0b_\mu b_\nu}2Ae^{ik_0|x|}\no&&-\frac{k_0^2b_\mu b_\nu}2\frac{B}{\sqrt F}e^{i\sqrt F|x|}.
\ea
Let us now contract the propagator with the initial and final polarization vectors
\ba
\epsilon^\mu\mathcal D_{\mu\nu}(k_0,x)\epsilon_0^\nu
&\approx&\frac{\vec\epsilon\cdot\vec\epsilon_0}{2k_0}e^{ik_0|x|}+
\frac{(\vec\epsilon\cdot\hat b)(\vec\epsilon_0\cdot\hat b)}{2k_0}e^{ik_0|x|}\cr&&-
\frac{(\vec\epsilon\cdot\hat b)(\vec\epsilon_0\cdot\hat b)}{2k_0}e^{ik_0|x|}e^{i\frac{\vec b^2}{2m^2}k_0|x|},
\ea
where $\hat b=\vec b/|\vec b|$.
Its squared modulus is
\ba\label{prob}
&&|\epsilon^\mu\mathcal D_{\mu\nu}(k_0,x)\epsilon_0^\nu|^2=\no
&&\frac1{4k_0^2}\left[E_1^2+4\left(E_1E_2+E_2^2\right)\sin^2\left(\frac{\vec b^2}{4m^2}k_0|x|\right)\right],
\ea
where
\be\label{e}
E_1=\vec\epsilon\cdot\vec\epsilon_0,\qquad E_2=(\vec\epsilon\cdot\hat b)(\vec\epsilon_0\cdot\hat b).
\ee
This quantity, once properly normalized, describes the quantum mechanical probability of measuring the polarization
represented by the vector $\epsilon$ at a distance $|x|$ from the origin, where it was created with a polarization represented
by $\epsilon_0$. Since we restrict ourselves to the case $\vec k\cdot\vec B=0$ and assume that
the polarization vectors are orthogonal to the direction of propagation, we can write
\be
\hat k=\hat x,\quad \hat b=\hat y,\quad\vec\epsilon=\cos\alpha\hat y+\sin\alpha\hat z,\quad \vec\epsilon_0=\cos\beta\hat y+\sin\beta\hat z,
\ee
so that
\be\label{angles}
E_1=\cos(\alpha-\beta),\quad E_2=\cos\alpha\cos\beta.
\ee
The extrema of \eqref{prob}, for a given initial angle $\beta$, are at
\be
\tan2\alpha(x)=\frac{[1+2f(x)]\sin2\beta}{4f(x)+[1+4f(x)]\cos2\beta},
\ee
where
\be
f(x)=\sin^2\left(\frac{\vec b^2}{4m^2}k_0|x|\right),
\ee
corresponding to the values of the angle where the probability of finding the polarization vector is maximum or minimum.
The mean value of the angle is
\be
\bar\alpha(x)=-\frac12\frac{\left[1+2f(x)\right]\sin2\beta}{[1+4f(x)]+4f(x)\cos2\beta}.
\ee
If the electric field is initially parallel to the magnetic field, it remains parallel, i.e. $\alpha(x)=0$. Otherwise
a rotation in the plane of polarization appears.

The parameter characterizing the evolution is $k_0 |x| \vec b^2 / 2m^2$. Usually\cite{raffelt} mixing is treated via
the classical evolution equation
\be
\left[\omega^2 + \partial_x^2 + \left(\begin{matrix}0    &     0     &     0 \cr
                                              0    &     0   &  \omega b \cr
                                              0    &     \omega b    &  -m^2 \end{matrix}\right)\right]
\left(\begin{matrix} \omega\epsilon_\perp \cr
               \omega\epsilon_\parallel \cr
                a    \end{matrix} \right)= 0.
\ee
Note that the contribution from the Euler-Heisenberg lagrangian
induced by the virtual contribution of electrons \cite{adler} has not been included.
It is not difficult to verify that both methods lead to the same dispersion relations in the case
where $\eta_0=0$.

\section{Influence of the axion background}
Now we return to the case with the axion background. In the limit where there is no magnetic field
we recover the pole structure already discussed in the first section,
$\omega_\pm^2= \vec k^2 \pm\vert \eta_0\vert \vert \vec k\vert$, whose implications in an
astrophysical context were discussed in \cite{sasha,cosmicrays,ourfirst}.

When the magnetic field is present, in addition to these poles
we have three additional poles, manifest in the third term in (\ref{propagator}),
that in the limit $\eta_0\to 0$ correspond to the ones described following equation (\ref{propmagnetic}).
Taking into account that $\eta_0$ is a very small quantity we shall disregard terms
proportional to $\eta_0^2$ in what follows. Then these three poles exactly agree with the ones discussed in the
previous section.

It is certainly of interest to consider how the simultaneous presence of the magnetic field and the
cold axion background influences the kinematics.
If we consider the processes $e\to e \gamma$ or $p \to p \gamma$ discussed in the
introduction, in the presence of both
the cold axion background and the magnetic field the kinematical restrictions
change. We defer this analysis to a separate publication.

In order to consider the rotation of the polarization plane we note once again that
$\eta_0$ is a rather small parameter. We shall
neglect in the propagator all terms quadratic in $\eta_0$; then \eqref{propagator} becomes
\ba
\mathcal D_{\mu\nu}(k)&=&\frac{-i g_{\mu\nu}}{k^2}+\frac{ik_0^2b_\mu b_\nu}{k^2[k^2(k^2-m^2)-k_0^2\vec b^2]}\cr&&-g_\mu^j g_\nu^l\frac{\hh k_0^2\left[b_j(\vec b\times \vec k)_l-b_l(\vec b\times \vec k)_j\right]}{k^4\left[k^2(k^2-m^2)-k_0^2\vec b^2 \right]}
\ea
Now we contract with the polarization vectors:
\ba
\epsilon^\mu\mathcal D_{\mu\nu}(k)\epsilon_0^\nu
&=&\frac{iE_1}{k^2}+\frac{iE_2k_0^2\vec b^2}{k^2[k^2(k^2-m^2)-k_0^2\vec b^2]}\cr&&-\frac{E_3\hh k_0^2\vec b^2k_1}{k^4\left[k^2(k^2-m^2)-k_0^2\vec b^2 \right]},
\ea
where $E_1$ and $E_2$ are given in \eqref{e}, \eqref{angles} and
\be
E_3=(\vec\epsilon\cdot\hat b)[\vec\epsilon_0\cdot(\hat b\times \hat k)]-(\vec\epsilon_0\cdot\hat b)[\vec\epsilon\cdot(\hat b\times \hat k)]=\sin(\alpha-\beta)
\ee

We implement for the piece proportional to $\eta_0$ a decomposition similar to the one described in the previous section.
\ba
&&\frac1{k^4\left[k^2(k^2-m^2)-k_0^2\vec b^2 \right]}=\no
&&~~\frac{\tilde A}{\vec k^2-k_0^2}+\frac{\tilde B}{\vec k^2-F}+\frac{\tilde C}{\vec k^2-G}+\frac{\tilde D}{(\vec k^2-k_0^2)^2}.
\ea$F$ and $G$ have been derived before. The new (tilded) coefficients are
\ba
\tilde B&=&\frac1{(k_0^2-F)^2(F-G)}\approx\frac{m^2}{k_0^4\vec b^4}\\
\tilde C&=&-\frac1{(k_0^2-G)(F-G)}\approx\frac1{m^6}\\
\tilde D&=&\frac1{(k_0^2-F)(k_0^2-G)}\approx-\frac1{k_0^2\vec b^2}\\
\tilde A&=&-\tilde B-\tilde C\approx-\frac{m^2}{k_0^4\vec b^4}.
\ea
We will again consider the propagation of an electromagnetic plane wave of frequency $\omega=k_0$ in the $\hat x$ direction, perpendicular
to the magnetic field. We have
\ba
&&\epsilon^\mu\mathcal D_{\mu\nu}(k_0,k_1)\epsilon_0^\nu=\cr
&&\frac{-iE_1}{k_1^2-k_0^2}+iE_2k_0^2\vec b^2\left(\frac{A}{k_1^2-k_0^2}+\frac{B}{k_1^2-F}+\frac{C}{k_1^2-G}\right)\no
&&-E_3\hh k_0^2\vec b^2 \left(\frac{\tilde Ak_1}{k_1^2-k_0^2}+\frac{\tilde Bk_1}{k_1^2-F}+\frac{\tilde Ck_1}{k_1^2-G}+\frac{\tilde Dk_1}{(k_1^2-k_0^2)^2}\right).\no
\ea
Then
\ba
&&\epsilon^\mu\mathcal D_{\mu\nu}(k_0,x)\epsilon_0^\nu
\approx\no
&&\frac{E_1}{2k_0}e^{ik_0|x|}+E_2\frac{e^{ik_0|x|}}{2k_0}\left(1-e^{i\frac{\vec b^2}{2m^2}k_0|x|}\right)\no
&&+iE_3\frac{\hh m^2}{2k_0^2\vec b^2}e^{ik_0|x|} \left(1-e^{i\frac{\vec b^2}{2m^2}k_0|x|}-i\frac{k_0|x|\vec b^2}{m^2}\right).
\ea
For small values of $ k_0|x|\vec b^2/m^2$ its square reduces to
\ba\label{probeta}
&&|\epsilon^\mu\mathcal D_{\mu\nu}(k_0,x)\epsilon_0^\nu|^2\approx\no
&&\displaystyle\frac1{4k_0^2}\left[E_1^2+\left(\frac{\vec b^2}{2m^2}k_0|x|\right)^2(E_1E_2+E_2^2)+3\eta_0|x|E_1E_3\right],\no
\ea
where terms of order $\eta_0^2$ have been neglected.

The extrema of \eqref{probeta} are at
\be
\tan2\alpha(x)=\frac{[1+2f(x)]\sin2\beta+3\eta_0|x|\cos2\beta}{4f(x)+[1+4f(x)]\cos2\beta-3\hh |x|\sin2\beta},
\ee
where
\be
f(x)=\frac{\vec b^4}{16m^4}k_0^2|x|^2,
\ee
and the mean value of the angle
\be
\bar\alpha=-\frac12\frac{\left[1+2f(x)\right]\sin2\beta+3\eta_0|x|\cos2\beta}{[1+4f(x)]+4f(x)\cos2\beta}
\ee
\section{Physical implications}
In this paper we have seen that high-energy charged particles moving in a spatially constant but
time varying axion background with  frequency $m$ radiate at a rate that agrees with the one computed
in \cite{cosmicrays,ourfirst}, where the approximation $\omega, k \gg m$ was assumed.
It was seen in \cite{cosmicrays,ourfirst} that the effect increases as the wave number
of the emitted photons decreases, and its possible detection
(if at all) is likely to occur in the MHz range of radiowaves. In this region $k\ll m$ and therefore the heuristic
arguments used in \cite{cosmicrays,ourfirst} could be questioned. The calculation presented here settles the issue.
The effect under discussion is quite real and to the best of our knowledge would constitute the clearest
(perhaps even the only\footnote{A possible exception would be the confirmation of \cite{sik} that axions or ALP form
Bose-Einstein condensates and caustics appear as a consequence in the galactic structures.}) observational evidence that
axions or ALP constitute the bulk of the dark matter component of the universe.
The so-called direct observation experiments, such as ADMX, CAST or analogous ones may find
evidence for the existence of a particle
with the properties of the axion or ALP but this would not prove (although it would certainly be a
tremendous hint) that axions or ALP constitute the missing mass of the universe.

It is of course unfortunate that the amount of
radiation predicted by the effect discussed here is very small; it is possibly within the
sensitivity of long-wave radio antennae
being commissioned or already deployed  but
around six orders of magnitude below the average value of the Galaxy synchrotron radiation background
for the best value of $\eta_0$.
The effect approximately scales as  $\displaystyle\eta_0\left(\eta_0/k\right)^{1.5}$.
We expect this parameter to be $\eta_0 < 10^{-20}$ eV
given the current bound for $f_a$ and the matter density (assumed to be due to cold axions).

In principle observations in regions of low magnetic field could increase the signal/background ratio by several orders
of magnitude as the synchrotron radiation is proportional to $\vec B^2$, assuming that the flux of cosmic rays stays at the
average value in galactic regions of low magnetic field. It should be noted that the assumption for the electron flux (electrons
radiate most in the present mechanism\cite{cosmicrays}) was taken very conservatively to be the value measured by satellites,
likely to be a gross underestimate of the value in inner parts of the Galaxy. On the contrary, the background quoted is the
observed value. In view of these considerations we believe that is important to refine the estimates before concluding whether
this axion-induced Bremsstrahlung could be measurable or not, or used to place relevant bounds for $\eta_0$ and hence
on $f_a$. Note that photons radiated via this effect are circularly polarized, while
synchrotron light is polarized in the plane of motion; measuring polarization may therefore help in the detection of the effect.

In any case it should be said that the effects under discussion could be considerably enhanced for ALP models (assuming that the
corresponding ALP particle condensates similarly to axions proper) because some constraints are evaded by these models.
This is certainly something to have in mind and worth of further investigation.

Another remarkable consequence of the presence of a slowly varying axion or ALP background is the fact that some wave-lengths
(actually very narrow bands, see figure 1 and its caption)
are forbidden in the universe (or at least where there are substantial concentrations of cold dark matter, if this
is constituted by cold axions or ALP). This opens of course a door for another line of experiments that could
potentially probe these forbidden wave-lengths. The viability of these experiments, which appear very difficult unless the axion
or ALP mass is known beforehand, deserves further investigation too.

We have also studied the effect on the polarization of photons propagating in this oscillatory pseudoscalar background. We assume
that $\eta(t)=\eta_0$ provides a good guidance. The results
presented here have to be considered as exploratory and a more detailed account will be presented elsewhere. The relevant
quantity that governs the change in the plane of polarization is the ratio $\omega x \vec b^2 / 2 m^2$. The value of
$\vert \vec b \vert$ ranges from $10^{-15}$ eV for magnetic fields of $10$ T (such as the ones employed in CAST)
to $10^{-6}$ eV for magnetar-strength fields, assuming that $f_a \sim 10^{7}$ GeV.
Taking $m\sim 0.1 $ eV as a reference value for the axion mass this corresponds to the following approximate range
\be
10^{-28} < \frac{\vec b^2}{m^2} < 10^{-10}.
\ee
We qualitatively reproduce previous results\cite{raffelt} in the case where only the magnetic field is considered.
However, since we have derived the complete quantum propagator when photons propagate through an oscillating
cold axion coherent background we can examine the modifications due to it. We find that, quite remarkably, the modification
in the ellipticity is independent of the light wavelength and also of the axion mass. It is probably even more notable that
it is also independent of the magnetic field itself, even if one needs to introduce one to begin with. We should warn the
reader that because the results on the polarization are derived for constant $\eta$ they are strictly valid for very short distances
($< 2\pi/m$) only. The astrophysical consequences of the results presented here are yet to be fully
explored. Note that to provide more realistic results other medium effects (such as an effective photon mass
or the Euler-Heisenberg effective lagrangian) should be considered too.

Clearly the presence of the cold axion background modifies the properties of photon propagation in rather interesting ways.
The modifications are tiny, but some of them may perhaps be experimentally or observationally explored. This could possibly
shed some light on the nature of dark matter.

\section*{Acknowledgements}
We acknowledge the financial support MICINN through projects FPA2010-20807, Consolider CPAN as well as
the support of SRG2009502.
We thank A. Andrianov and F. Mescia for numerous discussions. We would also like to thank K. Zioutas for
encouraging us to examine in detail the effect of magnetic fields.


\begin{thebibliography}{99}

\bibitem{PQ} R.D. Peccei, H.R. Quinn, Phys. Rev. Lett. 38 (1977) 1440;
 S. Weinberg, Phys. Rev. Lett. 40 (1978) 223;
 F. Wilzcek, Phys. Rev. Lett. 40 (1978) 279.

\bibitem{darkmatter} L.Abbott and P. Sikivie, Phys. Lett. 120B, 133 (1983); M. Kuster, G. Raffelt and B. Beltran (eds), Axions: Theory, Cosmology and Experimental Searches, Lecture Notes in Physics 741 (2008).

\bibitem{other} R. Battesti et al., Lect. Notes Phys. 741, 199 (2008); P. Sikivie, Phys. Rev. Lett. 51, 1415 (1983) [Erratum ibid.
52, 695 (1984)]; D. Dicus, E. Kolb, V. Teplitz and R. Wagoner, Phys. Rev. D 18, 1829 (1978); G. Raffelt,
Phys. Rev. D 33, 897 (1986); D. Lazarus et al. Phys. Rev. Lett. 69, 2333 (1992); Y. Inoue et al. Phys. Lett. B
668, 93 (2008); G. Raffelt, Lect. Notes Phys. 741, 51 (2008).

\bibitem{pdg} K. Nakamura et al. (Particle Data Group), J. Phys. G 37, 075021 (2010)

\bibitem{searches}  F. Bergsma et al. [CHARM Collaboration], Phys. Lett. B 157, 458 (1985)

\bibitem{mass}  J. Jaeckel, E. Masso, J. Redondo, A. Ringwald and F. Takahashi, Phys. Rev. D 75, 013004 (2007)

\bibitem{cast} E. Arik et al [CAST collaboration], JCAP02(2009)008.

\bibitem{models} M. Dine, W. Fischler and M. Srednicki, Phys. Lett. B, 104, 199 (1981); A.R. Zhitnitsky, Sov. J. Nucl. Phys. 31, 260 (1980); J. E. Kim, Phys. Rev. Lett. 43, 103 (1979); M. A. Shifman, A. I. Vainshtein and V. I. Zakharov, Nucl. Phys. B 166, 493 (1980).

\bibitem{cosmicrays} D. Espriu, F. Mescia, A. Renau, JCAP 1108:002  (2011).

\bibitem{ourfirst} A. Andrianov, D. Espriu, F. Mescia and A. Renau, Phys. Lett. B 684 (2010) 101.

\bibitem{sasha} A.A. Andrianov, D. Espriu, P. Giacconi and R. Soldati,  JHEP 0909:057,2009;

\bibitem{kp} R. de L. Kronig and W.G. Penney, Proc. Roy. Soc. London Ser. A 130, 499 (1931).

\bibitem{raffelt} G. Raffelt  and L. Stodolsky, Phys. Rev. D 37, 1237 (1988).

\bibitem{adler} S.L. Adler, Ann. Phys. 67 (1971) 599; Z. Bialynicka-Birula and I. Bialynicki-Birula, Phys. Rev. D 2 (1970) 2341;
J.S. Heyl and L. Hernquist, J. Phys. A 30 (1997) 6485; W. Dittrich and H. Gies, hep-ph/9806417; H. Gies, hep-ph/0010287;
V.A. De Lorenci, R. Klippert, M. Novello, J.M. Salim, Phys. Lett. B 482 (2000) 134.

\bibitem{sik} P. Sikivie, Phys. Lett. B567, 1 (2003).

\end{thebibliography}
\end{document}